\documentclass[10pt,twocolumn]{article}

\usepackage{graphicx}

\usepackage{dcolumn}
\usepackage{bm}
\usepackage{mathtools,lipsum,cuted}

\usepackage{subcaption}
\usepackage{physics}
\usepackage{makecell}
\usepackage{diagbox}
\usepackage{xifthen}
\usepackage{url}

\newcommand{\Q}[1][]{\mathcal{Q}^{#1}}

\newcommand{\mixing}[1][]{\ifthenelse{\isempty{#1}}{\vb*{\beta}}{\beta_{#1}}}
\newcommand{\problem}[1][]{\ifthenelse{\isempty{#1}}{\vb*{\gamma}}{\gamma_{#1}}}

\newcommand{\hamming}[2]{d_{#1#2}}
\newcommand{\mhamming}[3]{g_{#1#2#3}}

\newcommand{\Sq}{S_q}
\newcommand{\Sc}{S_c}

\begin{document}

\title{Entropic property of randomized QAOA circuits}

\author{%
Chernyavskiy A.Yu.$^{1}$, Bantysh B.I.$^{1}$, Bogdanov Yu.I.$^{1}$ \\
\textit{\small 1 -- Kurchatov Institute National Research Centre, 123098, Moscow, Russia}\\
}

\date{andrey.chernyavskiy@gmail.com}

\maketitle

\begin{abstract}
Quantum approximate optimization algorithm (QAOA) aims to solve discrete optimization problems by sampling bitstrings using a parameterized quantum circuit. The circuit parameters (angles) are optimized in the way that minimizes the cost Hamiltonian expectation value. Recently, general statistical properties of QAOA output probability distributions have begun to be studied. In contrast to the conventional approach, we analyse QAOA circuits with random angles. We provide analytical equations for probabilities and the numerical evidence that for unweighted Max-Cut problems on connected graphs such sampling always gives higher entropy of energy distribution than uniform random sampling of bitstrings. We also analyse the probability to obtain the global optima, which appears to be higher on average than for random sampling.  
\end{abstract}

\vspace{2pc}
\noindent{\it Keywords}: entropy, quantum algorithms, NISQ, QAOA

\section{Introduction}

Quantum computing has the potential to revolutionize various fields, including optimization, machine learning, chemistry and cryptography, solving matrix and systems of linear equations \cite{montanaro2016quantum,moll2018quantum,preskill2018quantum,biamonte2017quantum,mcardle2020quantum,lisnichenko2022protein,kiktenko2018quantum,yan2022factoring,grebnev2023pitfalls, harrow2009quantum, xu2022quantum}. One of the most promising quantum algorithms for optimization problems is Quantum Approximate Optimization Algorithm (QAOA), which was proposed by Farhi et al. in 2014 \cite{farhi2014quantum}. The hardware implementation of QAOA requires Ising-type interaction between qubits or its simulation via a set of the device native quantum gates. M\o{}lmer-S\o{}rensen (MS) entangling gate on ion-based platform allows one to implement native XX interaction between any pair of qubits with bichromatic laser field \cite{manovitz2017fast}. Moreover, one could implement a global entangling gate to make different pairs of qubits interact in parallel \cite{lu2019global}. The performance of QAOA in the case of noisy global MS gate has been recently studied in \cite{lotshaw2023modeling}.

The question of the practical advantage of QAOA is being widely studied by a variety of approaches, but is still open. Most of such analysis was conducted in terms of so-called approximation ratio (ratio of an average solution to a global optima). Being one of the analysis standards for classical deterministic discrete optimization algorithms \cite{williamson2011design}, in the case of QAOA, approximation ratio doesn't analyse the probabilities (and corresponding computational complexity) of obtaining an optimal or good (near-optimal) solutions \cite{lotshaw2022approximate}. Recent researches have shown that the energy distributions of the QAOA circuits outputs for Max-Cut problems are close to the Boltzman distribution, where the effective temperature depend on the minimal energy, problem size and the QAOA circuits depth \cite{diez2023quantum,lotshaw2022approximate}. The entropy values of these pseudo-Boltzman distributions turned out to be higher than for random states with the same energy. Another output distribution property being theoretically and experimentally studied in the QAOA context (for Grover Mixer-QAOA) is ``fair sampling'': uniformity of sampling among optimal solutions \cite{bartschi2020grover, pelofske2021sampling, golden2022fair}.  

One of the convenient methods of searching good QAOA angles is a classical-quantum hybrid approach, when classical optimization algorithms are being used to optimize expectation value estimated by a set of quantum circuit shots (executions and measurements). However, it is known that sophisticated classical optimization techniques can show results comparable to just random search (e.g. \cite{liashchynskyi2019grid, kuncheva1998nearest}). In the case of QAOA, such situation can, for example, arise, when classical optimization algorithm's work is strongly disturbed by statistical noise inducted by the limited number of measurements. Also, most of the effective optimization techniques starts from random points, so the analysis of the behaviour of QAOA with random angles is informative and can be used to form a benchmark baseline for the hybrid classical optimization usage in QAOA. We consider the random uniform choosing of QAOA parameters (\textit{random parameters QAOA}, \textit{rpQAOA}). The sampling consists of two consequent steps of randomness: sampling random QAOA circuit angles, sampling output bitstrings by measuring circuit output. 

We analyse energy distributions sampled by rpQAOA and present analytical equation for the single-depth case. Computation of such distributions for Max-Cut problems on all connected graphs with 4 to 9 vertices (273189 graphs, data obtained from \cite{graphs}) shows that for all that graphs the entropy of the energy distribution appeared to be \textit{always} higher than for uniform random sampling of bitstrings. Random graphs of higher dimensions also hold this property. 
Further, we analyse the increase in the rate of obtaining global optimum of rpQAOA versus random sampling for several sets of problems.     
Finally, we consider the examples of arbitrary problem Hamiltonians with non-degenerate energy levels, and two energy levels with non-degenerate ground-state.

\section{QAOA}

The intuition behind the algorithm comes from the trotterization of adiabatic quantum computation \cite{farhi2014quantum}, which was proven to be a universal quantum computation model \cite{albash2018adiabatic}. One encodes an $n$-bit optimization problem into the searching of a ground-state (or minimal energy) of a problem Hamiltonian $H_P$ and computes the following $n$-qubit quantum state:
\begin{equation}\label{eq:qaoa_circ}
    \ket{\mixing,\problem}=\qty[\prod_{j=1}^{p}{e^{-i \mixing[j] H_M}e^{-i \problem[j] H_P}}] \ket{+}^{\otimes n},
\end{equation}
where $\ket{+}$ is the $+1$ eigenstate of $\sigma_x$ Pauli matrix, $H_M=\sum_{k}{\sigma_x^{(k)}}$ is the so-called mixing Hamiltonian, $\mixing=\{\mixing[j]\}$ and $\problem=\{\problem[j]\}$ are circuit parameters (angles), and $p$ is the QAOA circuit depth.
Many practical discrete optimization problems can be formulated in terms of quadratic unconstrained binary optimization (QUBO), where the cost function to be minimized is represented as
\begin{equation}
    \label{eq:QUBO}
    C(z) = \sum\limits_{i<j}s_{ij}z_i z_j+\sum\limits_i s_{ii} z_i,
\end{equation}
where $z_i\in\{-1,1\}.$ The corresponding problem Ising Hamiltonian is then straightforwardly obtained from (\ref{eq:QUBO}) by treating every $z_i$ as Pauli-Z operator acting on the $i$-th qubit. Note, that \eqref{eq:QUBO} can be easily rewritten in variables $x_i=\frac{1+z_i}{2}\in\{0, 1\}$ with corresponding cost function $F(x)$. We use the term ``bitstring'' for both binary cases. In our analysis, we consider three types of problems. The first one is the Maximum Cut (Max-Cut), which stands for ``cutting'' graph edges into two subsets with maximal edges between subsets. Then, the problem Hamiltonian is (\ref{eq:QUBO}) with non-zero $s_{ij}=1$ for graph edges and no linear terms ($s_{ii}=0$). Other two  problems are random weighted and unweighted QUBO problems with density (rate of non-zero $s_{ij}$ elements) uniformly distributed in $[0.1, 0.9]$ and integer coefficients uniformly distributed in $[-3,3]$ (weighted QUBO) or always equal 1 (unweighted QUBO). Note that formally an arbitrary $n$-bit QUBO problem can be reduced to the Max-Cut with the size $n+1$ \cite{rehfeldt2023faster}, however, we separate Max-Cut and QUBO because of the different generation of random problems. The adiabatic approach underlying QAOA is also used to solve QUBO problems \cite{maletin2023possibility}.

As follows from the adiabatic theorem, $\ket{\mixing,\problem}$ can be arbitrarily close to the ground state of $H_{P}$ for $p\rightarrow\infty$ and some angles. QAOA aims to obtain the solution with limited $p$. Finding appropriate angles with finite $p$ is the main difficulty of obtaining QAOA advantage. The original work on QAOA \cite{farhi2014quantum} proposed an efficient method of computing the expectation value $E(\mixing,\problem) = \expval{H_P}{\mixing,\problem}$ on classical computer for Max-Cut on 3-regular graphs, so angles can be computed by classical optimization without using quantum computers. Later, most of research (e.g. \cite{guerreschi2019qaoa, zhou2020quantum, fernandez2022study}) have been directed to classical optimization of $E$ with its estimation by running quantum circuits (however, the analysis of computational complexity of optimization algorithms is strongly limited).  Another perspective way is using the angles transfer conjecture \cite{brandao2018fixed,galda2021transferability,akshay2021parameter,lotshaw2021empirical,lotshaw2022scaling,shaydulin2023parameter}, which stands that the same or close angles can be used for wide sets of problems. Below we analyze the uniform random choice of QAOA angles.

\section{Entropic property of random parameters QAOA}
Uniform random sampling of bitstrings $z$ of size $n$ gives the probability distribution $P_{rs}(z)=1/2^n$ with maximal entropy. However, considering distributions of objective values of $C(z)$ (energies of $H_P$) provides different picture, because the cost $C(z)$ can be equal for different $z.$ So, the random sampling probability distribution of costs $C(z)$ is $$P_{rs}(c)=\sum\limits_{C(z)=c}P_{rs}(z)=w_c/2^n,$$ 
where $c$ are \textit{different} values of $C(z)$ (the number of unique $c$ may be much less than $2^n$), and $w_c$ defines the \textit{weight} of $c$ (the number of bitstrings $z: C(z)=c$). Note that such sampling corresponds both classical uniform random sampling and quantum measuring of $|+\rangle^{\otimes n}, $ which may be treated as QAOA with $p=0$.

Analogously, binning over the same costs (energies) we consider cost distribution of QAOA sampling
$$P_{\mixing,\problem}(c)=\sum\limits_{C(z)=c}P_{\mixing,\problem}(z)=\sum\limits_{C(z)=c}|\langle \mixing,\problem|z\rangle|^2$$
The distribution of random parameters QAOA (rpQAOA) is
\begin{equation}
    P_{avg}(c)=\frac{1}{\abs{\mathcal{R}}}\int\limits_{\mathcal{R}}{P_{\mixing,\problem}(c)\dd \mixing \dd \problem}.
    \label{eq:pavg}
\end{equation}
where $\mathcal{R}$ is the parameter space domain. We consider the binary Shannon entropy of the cost (energy) distribution for random sampling \mbox{$\Sc=-\sum_c{P_{rs}(c) \log_2P_{rs}(c)}$} and for rpQAOA \mbox{$\Sq=-\sum_c{P_{avg}(c) \log_2P_{avg}(c)}$}, where sums are taken over unique cost values.

The \textit{main conjecture} of the work is that for single-depth rpQAOA for unweighted Max-Cut problems one has $\Sq>\Sc.$ 

To strictly verify this conjecture on large sets of graphs, we have derived the analytical expression for single-depth $P_{avg}(c).$ Let us consider an arbitrary binary cost function $F(x)$ and the corresponding problem Hamiltonian $H_P = \sum_x{F(x)\ketbra{x}}$ (here we do not restrict $F(x)$ to be a quadratic function). One can compute the following probability of obtaining the result $x$ in single-depth QAOA circuit (e.g. see \cite{hadfield2022analytical}):
\begin{strip}
\begin{equation}\label{eq:px_param}
    P_{\mixing_1,\problem_1}(x)
    = \frac{\cos[2n](\mixing[1])}{2^n}\sum_{yy^\prime}{[-i\tan(\mixing[1])]^{\hamming{x}{y}}[+i\tan(\mixing[1])]^{\hamming{x}{y^\prime}}e^{-i\problem[1] [C(y)-C(y^\prime)]}},
\end{equation}
\end{strip}
where $\hamming{x}{y}$ is the hamming distance between bitstrings $x$ and $y$. Taking the average over the conventional domain of circuit parameters ($[0, \pi]$ for $\mixing_1$ and $[0,2\pi]$ for $\problem_1$) and assuming that the objective function $F(x)$ is integer valued we obtained the cost sampling probabilities for single-depth rpQAOA
\begin{equation}\label{eq:pi_avg}
    P_{avg}(f) = \frac{w_f}{2^n} + 
    \!\!\!\!\!\!\!
    \sum_{
        \substack{
            x;\; y \neq y^\prime \\
            F(x)=f \\
            F(y)=F(y^\prime) \\
            \text{$\hamming{y}{y^\prime}$ is even}
        }
    }{
    \!\!\!\!\!\!
    (-1)^{\hamming{x}{y}+\mhamming{x}{y}{y^\prime}}
    \frac{C_{n}^{2n} C_{\mhamming{x}{y}{y^\prime}}^{n}}{2^{3n} C_{2\mhamming{x}{y}{y^\prime}}^{2n}}
    },
\end{equation}
where $\mhamming{x}{y}{y^\prime} = (\hamming{x}{y}+\hamming{x}{y^\prime})/2$, and $C_k^n$ is the binomial coefficient. Note that one should know or compute $F(x)$ for all the input bitstrings in order to use equations \eqref{eq:px_param} and \eqref{eq:pi_avg}.

Using effective software implementation of computing (\ref{eq:pi_avg}), the conjecture was verified on all connected graphs with the number of vertices $n$ from 4 (6 graphs) to 9 (261080 graphs), which provides the strong evidence. The details on statistics of $\Delta S = \Sq-\Sc$ on the considered set of problems is presented in Table \ref{tab:Entropy}. For higher dimensions we used numerical approximation of (\ref{eq:pavg}) by averaging over 200 random angles and have verified the conjecture on 1000 graphs for each problem size $n$ from 10 to 20. We also found counter-examples showing that the conjecture doesn't hold for general QUBO problems and weighted Max-Cut with coefficients from $\{1,2\}$. 

\begin{table}
\caption{Cost (energy) distribution entropy increase in rpQAOA output for Max-Cut problem on all connected graphs with 3 to 9 vertices.}
\label{tab:Entropy}

\resizebox{\linewidth}{!}{%

\begin{tabular}{|c|c|c|c|c|c|c|}
\hline
$n$&4&5&6&7&8&9\\
\hline
\#graphs&6&21&112&853&11117&261080\\
\hline
min $\Delta S$&0.046&0.020&0.064&0.049&0.056&0.074\\
\hline
avg  $\Delta S$&0.145&0.160&0.175&0.182&0.187&0.191\\
\hline
max  $\Delta S$&0.295&0.366&0.418&0.563&0.640&0.817\\
\hline
\end{tabular}}

\end{table}

\section{Performance analysis of random parameters QAOA}
A common way to analyze the QAOA performance is to use the approximation ratio. For some particular problems it's reliability was supported by the concentration effect: the small energy deviation around the expectation value \cite{farhi2014quantum,farhi2022quantum}. For a general probabilistic algorithm the approximation ratio might be an unreliable measure. Consider the sampling with $0.5$ probabilities of obtaining global minimum and maximum: approximation ratio of such hypothetical algorithm is just $0.5$, but practically it efficiently solves the problem. Thus we center on the probability $P_q$ to sample the cost function value $c$ using a quantum probabilistic algorithm. We compare it to the classical uniform random sampling using the quantum multiplying power (QMP):  $\Q_c=P_q(c)/P_{rs}(c)$. We denote QMP for rpQAOA computed in the global minimum $c=c_{min}$ as $\Q[avg]_{min}$. While rpQAOA gives noticable QMP (which will be discussed further in details), its approximation ratio is close to random sampling (computations on the considered sets of problems show about $10\%$ median decrease), what seems to be the consequence of Boltzmann-like distributions of Max-Cut costs.

\begin{figure*}
\includegraphics[width=\linewidth]{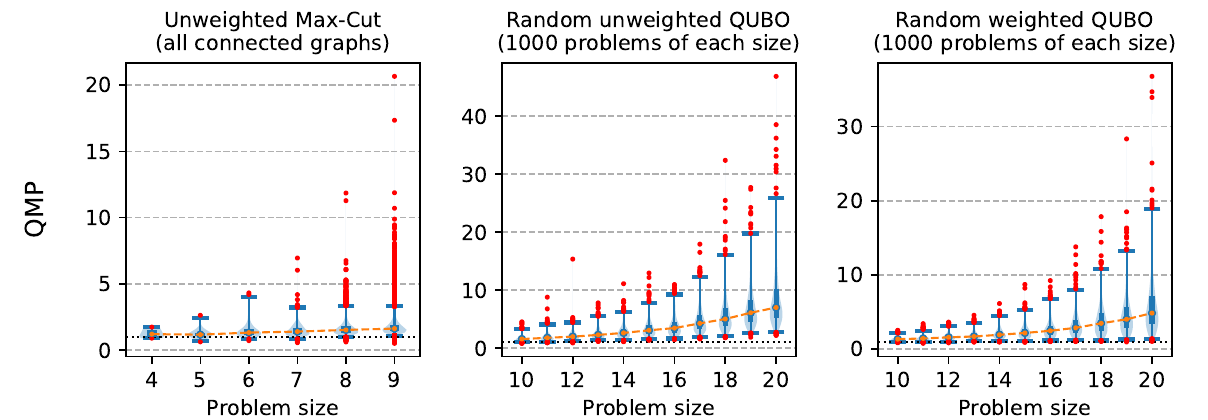}
\caption{Increase (in terms of QMP) of the global minimum probability of rpQAOA against random sampling: Max-Cut on all connected graphs, random unweighted QUBO problems (1000 problems of each size, density is random in [0.1,0.9]); weighted QUBO problems with integers weights in [-3, 3] (1000 problems of each size, density is random in $[0.1,0.9]$). Orange points show median values; filled bars show lower and upper quartiles; error bars show the 1st and 99th percentiles; red points are tails outside the 1st and 99th percentiles range; filled blue area is a so-called violin plot, which is similar to histogram, wider positions correspond to more values in a sample. Black dots are on the level of advantage $\Q[avg]_{min}=1.$}
\label{fig:qmp}
\end{figure*}

\begin{figure}
\includegraphics[width=1\linewidth]{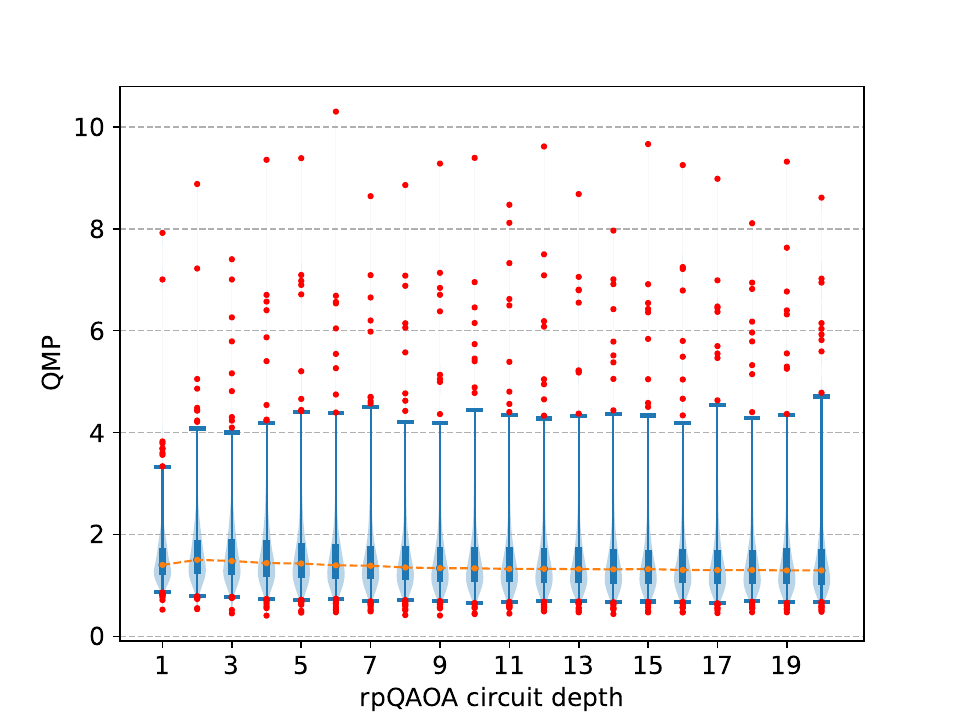}
\caption{The dependency of $\Q[avg]_{min}$ on the QAOA circuit depth $p$ on the set of all connected graphs with 7 vertices. The legend is the same with Fig. \ref{fig:qmp}}
\label{fig:qmp_vs_p}
\end{figure}

Using the distributions data obtained in the previous Section by semi-analytical and numerical computations of $P_{avg}(c),$ we have analysed QMP for all connected graphs with 4 to 9 vertices and for random graphs of larges sizes. Results are presented on Fig. \ref{fig:qmp}. The example of dependency of QMP on the depth $p$ is presented on Fig. \ref{fig:qmp_vs_p}, which shows that QMP is nearly independent of $p$ (the behaviour of the entropy increase $\Delta S$ is similar).  

Let's briefly analyze the computational complexity of rpQAOA. Fixing the desired probability $P_{goal}$ of obtaining the global minimum $c_{min}$ in a series of measurements, one can calculate the necessary number of measurements $\log_{1-P_{min}}{(1-P_{goal})}.$ We analyse the median value $T(n)$ (on the set of problems of the same size $n$) of $-1/\log{(1-P_{min})}, $ which corresponds to the computational complexity up to a constant. Considering that rpQAOA gives some polynomial speed-up, linear fit of $\log_2{T(n)}$ (which is close to linear for the considered data) gives approximately $T(n) = O(2^{0.86n})$ for Max-Cut, which predictably   is not better (even for median) than the best (to our knowledge) in terms of big $O$ classical algorithm having the complexity $O(m^32^{wn/3})$, where $m$ is the number of clauses (edges) and $w<2.376$ \cite{williams2005new}. We have also obtained $T(n) = O(2^{0.81n})$ for the random unweighed QUBO problems data.

Finally, we analyse two examples using (\ref{eq:pi_avg}). First, consider the extreme case, when all the objective values for all $2^n$ bitstrings are distinct (non-degenerate energy levels of $H_p$). One can easily see it results in $P_i^{avg}=1/2^n$, the maximal entropy 
$\Sc=\Sq=n$, and $\Q[avg]_{min}=1$, i.e. no advantage in using the rpQAOA over classical uniform random sampling.

Next, consider the cost function of the form $F(x_0)=f_{1}$, $F(x\neq x_0)=f_{2}$ with a single optimal answer $x_0$ (two energy levels of $H_p$, non-degenerate ground state). It gives the following equation for average rpQAOA probability to get $x_0$:
\begin{equation}
    P_{avg}(f_1) = \frac{1}{2^n} - \frac{1}{2^{2n-1}} + \frac{C_n^{2n}}{2^{3n-1}}.
\end{equation}
Both classical $H_c$ and quantum $H_q$ entropy drop exponentially with increasing problem size $n$, and $\Delta S \geq 0$ (since $\flatfrac{1}{2^n}\leq P_1^{avg} \leq \flatfrac12$ for any $n$). However, $\Q[avg]_{min}$ drops polynomially for $n \gg 1$:
\begin{equation}
    \Q[avg]_{min} \xrightarrow{n \gg 1} 1 + \frac{2}{\sqrt{\pi n}}.
\end{equation}

Note that both above examples cannot be represented in the unweighted QUBO form.

\section{Discussion}

Uniform random sampling of bitstrings is the natural efficiency baseline for classical optimization algorithms. In hybrid quantum-classical QAOA approach one can observe the absence of the optimization advantage over random angles by e.g. reducing the number of samples in expectation value estimation.
Surprisingly, we found out that random angles QAOA (rpQAOA) performance was higher on average than the random sampling for unweighted MaxCut problems. To our knowledge, the only previous work analysing random QAOA parameters was \cite{larkin2022evaluation}, which provided valuable detailed results of approximation ration on different sets of problems. However, as it was the main goal of the work, random parameters were just briefly analysed on a particular set of random 3-regular graphs and in another context.

The entropic conjecture formulated in this work may shed some light on the nature of the obtained increase in the probability of global minima. Another research with the detailed entropy analysis in the context of QAOA is the work of Lotshaw et al. \cite{lotshaw2022approximate}. Authors analysed output entropies of QAOA with optimized angles, and numerically obtained on the set of random examples that QAOA entropies are higher than entropies of specially generated random states with the same energies. Authors also found correlations of QAOA output behaviour with theoretic properties of Boltzmann entropy. In contrast, we analyse rpQAOA output and compare it to the corresponding classical uniform random sampling. It's important to emphasize that the obtained inequality was verified on \emph{all} $273189$ connected graphs with $4$ to $9$ vertices using the derived analytical equation, which provides a strong evidence. Results of \cite{lotshaw2021empirical} and ours appear to be the rare case, when entropy growth is useful.

\section{Conclusion}

In this work, we have performed the detailed analysis of the efficiency of QAOA with random parameters (rpQAOA). We have found that the probability to sample the global minimum of unweighted QUBO problems is higher for rpQAOA than for the uniform random sampling of bitstrings. During the investigation of this effect we have found the following property of the rpQAOA energy output distribution: for unweighted Max-Cut problems the distribution entropy is always higher than for the uniform random sampling. This is a rare example when the increasing entropy assists problem solving. The obtained results draws a benchmark baseline for the analysis of QAOA performance. The next reasonable step of rpQAOA research is expanding the range of the problems being analysed including Sherington-Kirpatrik spin glass model. 

\section{Acknowledgments}

This work was supported by the State Program no. FFNN-2022-0016.

\bibliographystyle{unsrt}
\bibliography{article.bib}

\end{document}